\documentclass[12pt, prd, showpacs]{revtex4}
\usepackage{amsmath}
\usepackage{amssymb}

\begin{document}

\title{Regular black holes and energy conditions}
\author{O. B. Zaslavskii}
\affiliation{Astronomical Institute of Kharkov V.N. Karazin National University, 35
Sumskaya St., Kharkov, 61022, Ukraine}
\email{ozaslav@kharkov.ua}

\begin{abstract}
We establish the relationship between the space-time structure of regular
spherically-symmetrical black holes and the character of violation of the
strong energy condition (SEC). It is shown that it is violated in any static
region under the event horizon in such a way that the Tolman mass is
negative there. In non-static regions there is constraint of another kind
which, for a perfect fluid, entails violation of the dominant energy
condition.
\end{abstract}

\maketitle




The famous theorem \cite{hp} establishes the connection between the validity
of strong energy condition (SEC) and the appearance of singularities inside
black holes. Correspondingly, if a black hole is regular, SEC\ is
necessarily violated somewhere inside the horizon. Examples of such black
holes were constructed \cite{bar}, \cite{dym}. Researches on regular black
holes are continuing to develop (see works \cite{kreg}, \cite{bu}, review 
\cite{a} and references therein) that makes it important to formulate more
precisely criteria, how to evaluate the degree of such violation. In the
present work we suggest such a very simple and clear criterion. It is
formulated in terms of the well-known quantity - Tolman mass \cite{tol} and,
thus, has an integral character. Additionally, our derivation can be
considered as a simplified version of the proof of the theorem \cite{hp} for
the particular situation of the spherically-symmetrical black holes.

We restrict ourselves by spherically-symmetrical black hole metrics (though
generalization to the distorted case is straightforward). For definiteness,
we choose the gauge in which the metric can be written as 
\begin{equation}
ds^{2}=-dt^{2}f+\frac{du^{2}}{f}+r^{2}(u)d\omega ^{2}\text{.}  \label{m}
\end{equation}

We assume that, in general, there are zeros of the function $f$ at $u=u_{1}$%
, $u=u_{2}$, ...$u=u_{N}\equiv u_{h}$ where roots are enumerated from the
left to the right. The outmost root viewed from the outside corresponds to
the black hole event horizon. We also assume that there is a regular center
in the system. This means that (i) there exist the point $r=0$ in the system
such that (ii) all curvature invariants are finite there. To achieve this
goal, the metric functions $f$ and $r$ should obey the conditions which in
the cooridantes (\ref{m}) read $f\left( \frac{dr}{du}\right) ^{2}=1+O(r^{2})$%
, $f=f(0)+O(r^{2})$ as $r\rightarrow 0$ (see e.g. \cite{bu}, \cite{s} or
references therein). The typical example is the (anti)de Sitter metric for
which $r=u$ and $f=1+const\ast r^{2}$.

We always can achieve $u=0$ by shifting to the appropriate constant, so that
the region $0\leq u<u_{1}$ is static, $f>0$ there. As the outer region, by
the definition, should be also static, the number $N$ of zeros (horizon) is
even. Then, it is clear that for non-degenerate roots 
\begin{equation}
f^{\prime }(u_{2k-1})<0,f^{\prime }(u_{2k})>0.  \label{der}
\end{equation}%
The adjacent roots can merge giving there $f^{\prime }(u_{2k-1})=f^{\prime
}(u_{2k})=0$. However, as this does not affect our consideration, for
simplicity of presentation we assume that roots are nondegenerate (unless
otherwise specified).

Now let us write down the expression for the $tt$ component of the Ricci
tensor for the metric which does not depend on $t$:

\begin{equation}
R_{t}^{t}=\frac{1}{\sqrt{-g}}\frac{\partial }{\partial x^{i}}(\sqrt{-g}%
g^{tt}\Gamma _{tt}^{i})
\end{equation}%
where we took into account from the very beginning that components $g_{ti}=0$
($i=r,\theta $,$\phi $). For the metric (\ref{m}) it gives us

\begin{equation}
-R_{t}^{t}r^{2}=\frac{1}{2}\frac{d}{du}(r^{2}f^{\prime })  \label{r00}
\end{equation}%
where $^{\prime }\equiv \frac{d}{du}$. Now we integrate this equality and
exploit the $tt$ components of the Einstein equations written in the form $%
R_{\mu }^{\nu }=8\pi (T_{\mu }^{\nu }-\frac{1}{2}T_{\alpha }^{\alpha }\delta
_{\mu }^{\nu })$. Here, we use the system of units with $G=c=1$. The
stress-energy tensor $T_{\mu }^{\nu }$ supporting the metric (\ref{m}) has
the form%
\begin{equation}
T_{\mu }^{\nu }=diag(-\rho ,p_{r},p_{\perp },p_{\perp })\text{.}
\end{equation}

Upon integration, we obtain immediately that%
\begin{equation}
4m_{T}(a,b)=(r^{2}f^{\prime })\mid _{a}^{b}  \label{eq}
\end{equation}

where, by definition, 
\begin{equation}
m_{T}(a,b)=4\pi \int_{a}^{b}dur^{2}(T_{k}^{k}-T_{0}^{0})\text{,}
\end{equation}%
is the Tolman mass calculated within the corresponding interval.

Let us consider the equality (\ref{eq}) in different intervals.

1) $a=0$, $b=u_{1}$. Thus, we take the interval between the center and the
first inner horizon. As $u=0$ corresponds to the center, $r(0)=0$. Further,
using eq. (\ref{der}) we immediately obtain that 
\begin{equation}
m_{T}(0,u_{1})<0\text{.}  \label{st0}
\end{equation}

2) $a=u_{2k}$, $b=u_{2k+1}$. This interval corresponds just to the static
region between two successive horizons. Then, $f^{\prime }(u_{2k})>0$, $%
f^{\prime }(u_{2k+1})<0$.\thinspace\ Again, we obtain%
\begin{equation}
m_{T}(u_{2k,}u_{2k+1})<0\text{.}  \label{st}
\end{equation}%
3) $a=u_{2k-1}$, $b=u_{2k}$. This interval corresponds to the non-static
region. In such a case $m_{T}$ does not have a direct physical meaning of
the mass but, for simplicity, we will still call it Tolman mass. Now,%
\begin{equation}
m_{T}(u_{2k-1,}u_{2k})>0.  \label{nst}
\end{equation}

As a result of properties (1) - (3), in the interval $u_{k}<u<u_{k+n}$
between corresponding horizons, the Tolman mass as the function of $u$ has $%
n-1$ zeros.

4) $a=r_{N}$, $b=\infty $. Assuming also that at infinity the metric is
asymptotically flat, we can write%
\begin{equation}
f=1-\frac{2m}{r}+o(\frac{1}{r})\text{, }r\approx u
\end{equation}%
where $m$ is the Schwarzschild mass.

Then, we obtain the standard mass formula \cite{mass}%
\begin{equation}
m=m_{T}+\frac{\kappa A}{4\pi }
\end{equation}%
where $m_{T}$ is the total Tolman mass in the outer region, $A=4\pi
r_{h}^{2} $ is the area of the event horizon, $\kappa =\frac{f^{\prime
}(u_{h})}{2}$ is the surface gravity.

The strong energy condition (SEC) requires that $Y\equiv \rho
+p_{r}+2p_{\perp }\geq 0$. Meanwhile, in the static region $%
T_{k}^{k}-T_{t}^{t}=Y$ and we see from (\ref{st0}), (\ref{st}) that in each
of static regions in our system $\int dur^{2}Y<0$, except from the outer
one. Thus, SEC is necessarily violated somewhere in each such a region and,
moreover, this violation is so strong that it amounts to the negativity of
the corresponding Tolman mass. If, by contrary, $Y>0$ everywhere in the
static region, the solution with a regular centre and a horizon is
impossible in accordance with the general theorem \cite{hp} and its
manifestation in the context of spherically-symmetrical space-times \cite{s}%
. Actually, for such space-times, we generalized the aforementioned
observation of \cite{s} and related the violation of SEC and its degree to
the space-time structure having, in general, an arbitrary number of horizons.

We also formulated the energy condition to be satisfied in the non-static
region that does not have meaning of SEC. Indeed, in the non-static regions
the above formulas written in terms of $T_{\mu }^{\nu }$ are still valid but
the meaning of separate components of $T_{\mu }^{\nu }$ changes. Now, in the
region where $f<0$ the coordinate $u$ is time-like, $u\equiv T$. In a
similar way, the coordinate $t$ is space-like, $t\equiv x$. Correspondingly, 
$T_{u}^{u}=T_{T}^{T}=-\rho $, $T_{t}^{t}=T_{X}^{X}=p_{x}$. Then, the
condition (\ref{nst}) takes the form

\begin{equation}
m_{T}(u_{2k-1,}u_{2k})=4\pi \int_{u_{2k-1}}^{u_{2k}}dur^{2}(2p_{\perp
}-p_{x}-\rho )>0\text{.}  \label{mnst}
\end{equation}

It cannot be in general related to the standard energy conditions directly.
However, in the isotropic case $p_{x}=p_{\perp }\equiv p$ that corresponds
to the perfect fluid we have%
\begin{equation}
m_{T}(u_{2k-1,}u_{2k})=4\pi \int_{u_{2k-1,}}^{u_{2k}}dur^{2}(p-\rho )>0
\end{equation}%
and we obtain the violation of the dominant energy condition (DEC) $p\leq
\rho $. It follows that the perfect fluid obeying DEC everywhere cannot be a
source for black holes with a regular centre.

In the highly anisotropic case, when $\left\vert p_{\perp }\right\vert \ll
\left\vert p_{x}\right\vert $, it is seen from (\ref{mnst}) that the
condition $p_{x}+\rho \geq 0$ cannot be satisfied everywhere in a given
non-static region, so WEC (weak energy condition), DEC and SEC are violated.
In the opposite case, $\left\vert p_{x}\right\vert \ll \left\vert p_{\perp
}\right\vert $, there is no definite constraint of this kind. For example,
if $\frac{\rho }{2}<p_{\perp }<\rho $, DEC is satisfied.

As is mentioned in introductory paragraphs, if we want to have a regular
black hole, SEC should be violated somewhere. More precisely, in our context
only the sigh of $Y$ is important relevant whereas two other manifestations
of SEC $p_{r}+\rho \geq 0$, $p_{\perp }+\rho \geq 0$ are irrelevant in
accordance with the observation made in \cite{s}. One may ask a question -
is it possible to have SEC violated everywhere due to $Y<0$? In the
intermediate non-static region the condition (\ref{mnst})\ should be
satisfied. In combination with $Y<0$, this would give us in some subregion
of such a region the conditions $p_{x}+\rho <2p_{\perp }<-p_{x}-\rho $, so
that WEC is to be also violated, $p_{x}+\rho <0$ and the presence of the
region with phantom matter is unavoidable. The sign of $p_{\perp }+\rho $ is
not now fixed in the situation under discussion.

The fact that the explicit form of the energy condition in non-static
regions changed as compared to the static ones is quite natural. The SEC is
obtained from the requirement $R_{\mu \nu }u^{\mu }u^{\nu }\geq 0$ where $%
u^{\mu }$ is a time-like vector. In the static region it is convenient to
take $u^{\mu }$ to be the four-velocity of an observer at rest, whence the
explicit expression for $Y$ is obtained. But in the non-static region, we,
as a matter of fact, integrated the same inequality but with a space-like
four-vector $u^{\mu }$.

It is instructive to compare our approach with that in the recent paper \cite%
{l} where the idea of replacing the central singularity by the de Sitter
core \cite{dym} was again discussed. The model comprising the vacuum
Schwarzschild and de Sitter region separated by smooth distribution of
matter was considered there. It was noticed in \cite{l} that on the border
between vacuum and matter under the Schwarzschild horizon DEC is violated.
Actually, this observation can be understood as follows. If some
configuration is smooth (no shells are present), in the static region this
entails, as is known, the continuity of the radial pressure. In the
non-static region for the metric of the type (\ref{m}) the radial pressure
and energy density mutually exchange their roles as is explained above.
Correspondingly, this requires the continuity of $\rho $. As in the vacuum
region $\rho =0$, the same holds on the boundary from the inner side and, as
result, DEC is necessarily violated for any $p_{x}\neq 0$. For the metric of
the type (\ref{m}) with $f<0$, this circumstance is independent of whether
or not the horizon is present. Meanwhile, the condition (\ref{mnst}) relies
on the presence of the horizon. In doing so, even if DEC is violated, WEC
and SEC can be satisfied.

The special case arises if two successive roots are degenerate. Then, both
in the static or non-static situation the corresponding quantity $m_{T}=0$.
If, in addition, the vacuum-like equation of state $p_{x}+\rho =0$ holds,
DEC can be satisfied on the border but SEC\ is violated somewhere since $%
p_{\perp }$ should alter its sign to ensure the zero Tolman mass, so $Y$
should change sign also.

We see that in regular black holes the violation of the strong energy
condition (unavoidable due to the singularity theorem) received simple
formulation in terms of the Tolman mass. This violation occurs just in
static regions and the natural measure of its degree is the Tolman mass
calculated between two adjacent horizons or between the centre and the first
horizon. In the non-static regions, the standard energy conditions (not only
SEC) can be either violated or satisfied. For a perfect fluid DEC\ is
violated.

It is worth noting that, instead of integrating eq. (\ref{r00}) between
horizons, we could do it in the intervals between successive minima and
maxima of $f(u)$. Then, as $f^{\prime }=0$ at limits of the integral, each
of such integrals is equal to zero, so contributions of regions in which SEC
is violated and those where (\ref{mnst}) is satisfied, exactly compensate
each other, thus giving another measure of violation of SEC.

Up to now, we discussed the regular black hole having a centre. Meanwhile,
it turns out that the class of such space-time is more diverse and can
include, in particular, models with $r\rightarrow r_{0}$ $\ $or $%
r\rightarrow \infty $ inside the horizon \cite{kreg}, \cite{bu}. In such a
case the condition (\ref{st0}) loses its meaning but other conditions retain
their validity.

To summarize, we found the direct and simple relations between the
space-time structure of regular black holes and energy conditions which must
be violated or satisfied and having different meaning in static and
non-static regions. In particular, the integral measure of violation of SEC
is established in terms of the Tolman mass. We would like to stress that it
is this kind of mass which turned out to be relevant for this purpose,
whereas, say, the Hernandez-Misner-Sharp mass \cite{mass2} defined according
to $\frac{2m}{r}=1-\left( \nabla r\right) ^{2}=1-$ $fr^{\prime 2}$ would not
give such information reducing simply to $m=\frac{r}{2}$ at each root of $f$.

It is of interest to extend the results of the present paper to the case of
rotating regular black holes. The separate issue is application of the
approach developed in the present paper to wormhole physics where
alternation of regions where the standard energy conditions are satisfied or
violated can also occur \cite{ks}.

\begin{acknowledgments}
This work was conceived during the conference QFEXT09 in Norman. I thank
organizers and especially Kim Milton for creation of so stimulating
atmosphere and Alexander Burinskii for interest and discussions. I am also
grateful to Sergey Sushkov for careful reading the manuscript and Kirill
Bronnikov for valuable comments.
\end{acknowledgments}

\end{document}